\documentclass[10pt,aps,pre,twocolumn,superscriptaddress,nobibnotes,nodoi,noeprint]{revtex4-2}

\usepackage{graphicx}
\usepackage{dblfloatfix}
\usepackage{amsmath,physics,bm,float}
\usepackage{soul}
\usepackage[colorlinks=true,citecolor=blue,urlcolor=blue,linkcolor=blue]{hyperref}
\usepackage{amssymb}
\usepackage{natbib}
\usepackage{upgreek}
\usepackage{mathtools}
\usepackage{appendix}
\usepackage{color,xcolor}

\usepackage[mathlines]{lineno}
\let\oldequation\equation\let\oldendequation\endequation
\renewenvironment{equation}{\linenomathNonumbers\oldequation}{\oldendequation\endlinenomath}
\let\oldalign\align\let\oldendalign\endalign
\renewenvironment{align}{\linenomathNonumbers\oldalign}{\oldendalign\endlinenomath}

\DeclareMathOperator*{\MSD}{MSD}

\newcommand{\hhu}{Institut f\"ur Theoretische Physik II: Weiche Materie, Heinrich-Heine-Universit\"at D\"usseldorf, Universit\"atsstra{\ss}e 1, D-40225 D\"usseldorf,
	Germany}
\newcommand{\tud}{Institut f\"ur Physik der kondensierten Materie, Technische Universit\"at Darmstadt, Hochschulstra{\ss}e 8, D-64289 Darmstadt, Germany}
\newcommand{\jgu}{Institut f\"ur Physik, Johannes Gutenberg-Universit\"at Mainz, 
	55128 Mainz, 
	Germany}
\newcommand{\sau}{Theoretische Physik, Universit\"at des Saarlandes, Campus E26, D-66123 Saarbr\"ucken, Germany
}

\begin{document}
	
	\title{Modeling dissipation in quantum active matter}
	
	\author{Alexander P.\ Antonov}
	\email{alexander.antonov@hhu.de}
	\affiliation{\hhu}
	
	\author{Sangyun Lee}
	\affiliation{\jgu}
	
	\author{Benno Liebchen}
	\affiliation{\tud}
	
	\author{Hartmut L\"owen}
	\affiliation{\hhu}
	
	\author{Jannis Melles}
	\affiliation{\hhu}
	
	\author{Giovanna Morigi}
	\affiliation{\sau}
	
	\author{Yehor Tuchkov}
	\affiliation{\jgu}
	
	\author{Michael te Vrugt}
	\email{tevrugtm@uni-mainz.de}
	\affiliation{\jgu}

	\date{\today}
	\begin{abstract}
		Active matter is characterized by a constant influx and dissipation of energy that gives rise to directed motion. Dissipation requires interactions with an external environment, such that extending the paradigm of active matter to a quantum framework requires an appropriate description of this environment. In this work, we consider a driven quantum particle undergoing noise and dissipation, with external driving exhibiting characteristics of classical activity. We model the non-unitary dynamics with time-local master equations and analyze the particle motion at different time scales for different forms of the master equations, satisfying different criteria. We systematically compare predictions on the dynamics of particle trajectories and thereby we uncover how the particle motion evolves under the interplay of quantum effects, dissipation, and active-like dynamics. These results are essential for guiding possible experiments aimed at realizing quantum analogues of classical active systems.
	\end{abstract}
	
	\maketitle
	
	\section{Introduction}
	
\begin{figure}[ht!]
		\includegraphics[width=0.6\columnwidth]{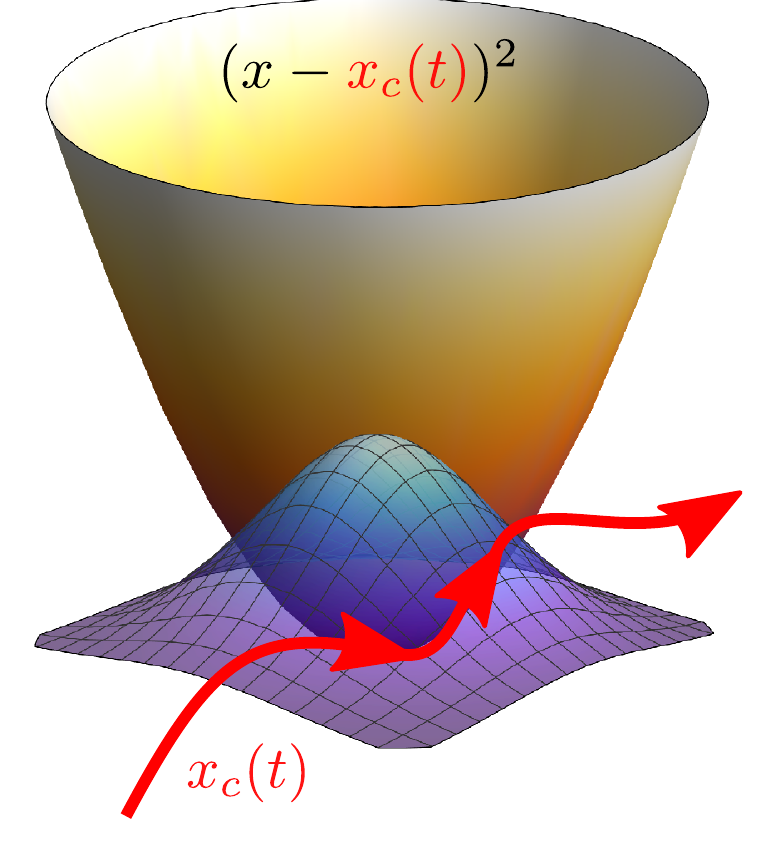}
		\caption{Schematic representation of a possible experiment -- an external parabolic potential (yellow) captures a quantum particle (blue wave function). The potential moves along a trajectory $x_c(t)$ corresponding to colored noise. Since the quantum particle is constantly driven away from the instantaneous trap center, 
			the system never truly reaches equilibrium, and the quantum particle thereby mimics the properties of active motion incorporated into the driving protocol $x_c(t)$.
		}
		\label{fig:setup}
	\end{figure}

	Quantum active matter is an emerging and rapidly developing field \cite{Adachi/etal:2022, Khasseh/etal:2023, Yamagashi/etal:2024, Takasan/etal:2024, Antonov/etal:2025, Nadolny/etal:2025, Penner/etal:2025} that aims at identifying quantum mechanical dynamics
	which can be classified as active \cite{Vrugt/etal:2025}. In classical physics, activity arises from the local absorption of energy from internal or external sources and its subsequent conversion into persistent, directed motion \cite{Marchetti/etal:2013, Elgeti/etal:2015, Bechinger/etal:2016}. Recent works have shown that the concept of active motion can be transferred to quantum systems \cite{Khasseh/etal:2023, Yamagashi/etal:2024, Takasan/etal:2024, Penner/etal:2025}. A paradigmatic example is a quantum particle bound to a potential, whose center undergoes stochastic fluctuation~\cite{Antonov/etal:2025}, see Fig.~\ref{fig:setup}. In this system, activity is introduced externally through the prescribed classic stochastic trajectory $x_c(t)$. The resulting quantum motion can then mimic key characteristics of the motion of ``self-propelled'' particles \cite{Vrugt/etal:2025, Marchetti/etal:2013, Bechinger/etal:2016}, i.e., ballistic motion at intermediate times and active diffusion at late times, while the system remains unbiased (isotropic) after averaging over many realizations. While the approach suggested in \cite{Antonov/etal:2025} generally works both in Hamiltonian and in dissipative systems, the latter case shows closer analogies with classical active matter and is therefore of particular importance. This brings to the fore the need to understand how to consistently model dissipation in activated quantum systems, which motivates the present study.
	
	In classical active matter, dissipation is a crucial factor since active matter dynamics and thermodynamics typically arise from the energy exchange with an external environment \cite{Das/etal:2020, Martinez/etal:2021, Fodor/etal:2022, Bebon/etal:2025}. A quantum mechanical analogue of active matter shall also possess these features and necessarily calls for an adequate formalism. However, the interaction of a quantum system with an external environment typically tends to induce loss of quantum coherence. Hence, one key question is to understand the role that a reservoir plays for quantum activity.
	
	The theory of open quantum systems \cite{breuer2002theory} provides a sound description of the dynamics of a quantum system that exchanges energy and establishes correlations with an external reservoir. In the presence of interactions with an external reservoir, the evolution of the quantum state is determined by a completely positive and trace-preserving (CPTP) time evolution of the system's density matrix. The latter describes the dynamics of the quantum probability amplitudes, from which interferences originate, and of the classical probabilities due to partial knowledge about the system, which determine the dynamics in the classical limit.
	
	Generators of the CPTP map can be cast in the form of Nakajima-Zwanzig master equations \cite{Nakajima1958,Zwanzig1960,teVrugtW2020}, which can be rigorously derived from a Hamiltonian description of coupled system and external environment by eliminating the environment's variables from the equations of motion of the system's variables \cite{breuer2002theory}. They are generally inhomogeneous differential equations for the system's density operator that are nonlocal in time. Their specific form depends on the initial state of the reservoir, on the initial correlations between system and reservoir, and on the system-reservoir interactions. These equations are tractable in few limits \cite{Hu/etal:1992} and are usually simplified by means of plausible approximations \cite{Hartmann:2020}. One prominent approximation is the Born-Markov one, for which the master equation becomes a homogeneous differential equation that is local in time \cite{breuer2002theory,
		Rivas/Huelga:2012,Hartmann:2020} (see also \cite{deVega/Alonso:2017}). Different forms of time-local master equation have been discussed and used in the literature. Some 
	satisfy the CPTP requirement \cite{Lindblad:1976, Carmichael:1993, Nathan/Rudner:2020}. For instance, Born-Markov master equations derived using the secular approximation \cite{Barnet/Knight:1986, breuer2002theory, stefanini2025lindblad} and the rotating-wave approximation \cite{Ford/Oconnell:1997, stefanini2025lindblad} are well suited for the description of quantum optical systems. Other approaches are designed to be thermodynamically consistent, making them more natural from the perspective of statistical mechanics \cite{Agarwal:1971,Soret/etal:2022, Wang/etal:2023b, Hewgill/etal:2021}. For an overview of time-local master equations and an accuracy assessment see \cite{Hartmann:2020}. This distinction is not merely theoretical. Different master equations can best describe different regimes and experimental scenarios \cite{Myatt:2000, Rivas/Huelga:2012}. Understanding how quantum active dynamics can emerge for different models, describing dissipation and dephasing, can be crucial for experiments aiming at observing quantum active dynamics.
	
	The purpose of this work is to analyze the effect of different types of quantum time-local master equations on the dynamics of quantum active matter. Our starting point is the framework of Ref.\,\cite{Antonov/etal:2025}, where the features of active matter reproduced by a quantum particle confined by a potential and driven with colored noise. In this paper we extend this approach to consider different generators of the open-quantum system dynamics. By systematically comparing several dissipative master equations, we uncover the interplay between 
	quantum effects and active-like dynamics across weak and strong dissipation regimes. Depending on the underlying dissipator, describing the non-unitary part of the dynamics, one may obtain a quantum-active evolution that ensures positivity of the density matrix, or one that reproduces the standard active matter dynamics in the classical limit -- but not both simultaneously.
	
	The paper is organized as follows: in Sec.~\ref{sec:model} we introduce the models and discuss the details of the numerical procedure. In Sec.~\ref{sec:results} we provide the results for the different master equations and discuss them. In Sec.~\ref{sec:conclusion} we draw our conclusions. Details of theoretical models and analysis are provided in the Appendices.
	
	\section{Open quantum system description of a quantum active particle}
	\label{sec:model}
	
	The purpose of this section is to introduce models of a quantum active particle in the presence of an external environment.
	
	\subsection{Model}
	
	Our starting point is the minimal model of quantum active matter of Ref.~\cite{Antonov/etal:2025}, which is inspired by the active Ornstein-Uhlenbeck particle (AOUP) and is one-dimensional. 
	In the classical AOUP model, the particle’s self-propulsion arises from colored noise $x_c$, which drives persistent active motion \cite{Szamel/etal:2014, Maggi/etal:2014, Caprini/Marconi:2018, Fily:2019, Martin/etal:2021, Keta/etal:2022, Schuettler/etal:2025}. Specifically, the variable $x_c(t)$ is the classical trajectory of a particle that solves the equation of motion
	\begin{equation}
		\tau\ddot{x}_c(t) = -\dot{x}_c(t) + \sqrt{2D}\eta(t),
		\label{eq:AOUP}
	\end{equation}
	where $\tau$ is the inverse damping rate (or equivalently, the persistence time), $D$ is the diffusion coefficient in units of the damping rate, and $\eta(t)$ is a Gaussian white noise with zero mean and correlations $\overline{\eta(t')\eta(t'')} = \delta(t'-t'')$. 
	
	In the quantum-mechanical model proposed in Ref.\ \cite{Antonov/etal:2025}, a quantum particle of mass $m$ is confined by a time-dependent harmonic potential. The potential has a fixed frequency $\omega$ and the center follows the prescribed Ornstein-Uhlenbeck trajectory $x_c(t)$. Thereby, the quantum particle is governed by a time-dependent Hamiltonian of the form
	\begin{equation}
		\hat{H}(x_c(t))=\frac{\hat{p}^2}{2 m} +\frac{1}{2} m \omega^2 \left(\hat{x}-x_c(t)\right)^2\,,
		\label{eq:hamiltonian}
	\end{equation}
	where $x_c(t)$ is a classical variable undergoing the dynamics of Eq.\ \eqref{eq:AOUP}; 
	$\hat{x}$ and  $\hat{p}$ are the canonically-conjugated position and momentum along $x$, satisfying the commutation relations $[\hat x,\hat p]=i\hbar$, where $\hbar$ is Planck's constant. The classical particle establishes a contact between the quantum particle and a thermal bath. 
	Since the external drive $x_c(t)$ is governed by the classical equation of motion \eqref{eq:AOUP} and unaffected by the state of the quantum particle, we can represent the state of the quantum system by a probability distribution over quantum trajectories $|\psi(x_c(t),t)\rangle$. Here, $|\psi(x_c(t),t)\rangle$ is the solution of the Schr\"odinger equation $\partial_t|\psi(x_c(t),t)\rangle=\hat{H}(x_c(t))|\psi(x_c(t),t)\rangle$ for a given trajectory $x_c(t)$.
	The state of the quantum particle is given by the density matrix $\hat\varrho(t)$ 
	\begin{eqnarray}
		\hat\varrho(t)=\int Dx_c(t) \mathcal{P}[x_c(t)]|\psi(x_c(t),t)\rangle\langle \psi(x_c(t),t)|\,,   
	\end{eqnarray}   
	where $\mathcal{P}[x_c]$ denotes the probability functional that the particle follows the trajectory $x_c(t)$ and the functional integral is performed over all paths,
	such that $\int Dx_c(t) \mathcal{P}[x_c(t)] = 1$.
	
	In this work, we assume that the quantum active particle is additionally coupled to a reservoir. The motion results from the interplay of the energy exchange of the quantum particle with an external reservoir and the coupling with the classical active particle, undergoing the dynamics governed by Eq.\ \eqref{eq:AOUP}. For a given trajectory $x_c(t)$, the state of the system is now described by a density operator $\hat{\rho}(x_c(t),t)$. 
	Since $x_c(t)$ represents the stochastic trajectory of a classical degree of freedom, physical observables must be averaged not only over quantum states but also over the ensemble of possible realizations of $x_c(t)$. In other words, the quantum system is subject to a noisy classical driving, and its effective behavior emerges only after performing both the quantum and the classical ensemble average.
	In general, the state  $\hat{\rho}(x_c(t),t)$ is evolved from the initial state $\hat{\rho}(x_c(0),0)$ by the CPTP map $\hat{\rho}(x_c(t),t)=\hat\Lambda_{x_c(t)}[\hat\rho(x_c(0),0)]$. By averaging over all trajectories, the state of the system is now given by 
	\begin{eqnarray}
		\label{eq:varrho}
		\hat\varrho(t)=\int Dx_c \mathcal{P}[x_c]\hat\rho(x_c(t),t)\,.   
	\end{eqnarray}   
	
	In order to determine the dissipative dynamics, we consider generators that are local in time, such that 
	$\partial_t\hat\rho={\mathcal L}(x_c(t),t)\hat\rho$. The superoperator ${\mathcal L}$ has the form
	\begin{eqnarray}
		& &{\mathcal L}(x_c(t),t)\hat\rho \\
		&=&-\frac{i}{\hbar}\left[ \hat{H}(x_c(t)), \hat{\rho}(x_c(t),t)\right]+{\mathcal{D}}(x_c(t), \hat{\rho}(x_c(t),t)),  \nonumber
		\label{eq:evolution}
	\end{eqnarray}	
	where $\left[\cdot, \cdot\right]$ is the commutator $\left[\hat{A}, \hat{B} \right] = \hat{A}\hat{B}-\hat{B}\hat{A}$ and ${\mathcal{D}}$, which describes the incoherent dynamics, is in what follows denoted by ``dissipator''.
	In this work, we assume time-local quantum master equation, which can be found as limiting case of a Nakajima-Zwanzig master equation when the environment’s correlation time is much shorter than the system’s characteristic time scale~\cite{breuer2002theory}. Being an approximation, it is not warranted that the corresponding map is CPTP. Positivity is warranted by the Lindblad form, as the one in the model of Ref.\ \cite{Antonov/etal:2025}. However, the Lindblad master equation does not deliver the correct asymptotic limit \cite{Ford/Oconnell:1997,Ingold_2002}. Other known forms satisfy the fluctuation-dissipation theorem but do not generally guarantee positivity \cite{Agarwal:1970,Caldeira/Legget:1983}. In what follows, we investigate how the physical behavior of the quantum active particle depends on these different choices.
	
	\subsection{Models for the dissipator}
	
	Here, we first introduce the different kinds of dissipators considered in this work, highlighting their defining features and the motivation behind their choice. Each dissipator provides an approximate description of different types of noise and of the regimes, possibly reflecting a range of experimental situations and setups. In what follows, we use the notation $\hat{\rho}(x_c(t), t) \equiv \hat{\rho}(t)$ for a single stochastic realization $x_c(t)$.
	\subsubsection{Lindblad dissipator}
	
	We consider a form of Lindblad dissipator, inspired by the one used in the model of a quantum active particle of Ref.\,\cite{Antonov/etal:2025}.\ The idea of \cite{Antonov/etal:2025} was to mimic active motion of a quantum particle by means of a trapping potential whose center follows the noisy trajectory of a classical particle. The quantum particle is also exposed to a thermal bath, whose effect is a weak perturbation with respect to the noise mediated by the classical particle. One issue of the form of Lindblad dissipator of \cite{Antonov/etal:2025} is that it does not preserve physical properties at large thermalization rates, see the discussion in Appendix \ref{sec:form}. In this work we consider a modification that accounts for motion of the trap along the trajectory $x_c(t)$. Before we give its explicit form, we introduce the annihilation and creation operators of the harmonic oscillator, $\hat{\tilde{a}}(t)$ and $\hat{\tilde{a}}^\dagger(t)$, such that Hamiltonian \eqref{eq:hamiltonian} takes the form $H=\hbar\omega (\hat{\tilde{a}}^\dagger(t)\hat{\tilde{a}}(t)+1/2$). These operators are related to the position and momentum operators of the oscillator centered at the origin, $\hat x=x_*(\hat a+\hat a^\dagger)$ and $\hat p=-i(\hat a-\hat a^\dagger)\hbar/(2x_*)$ with $x_*=\sqrt{\hbar/(2m\omega)}$, by the relations
	\begin{subequations}
		\label{eq:creation-annihilation}
		\begin{eqnarray}
			\hat{\tilde{a}}(t) & = & \sqrt{\frac{m\omega}{2\hbar}} \left( \hat{x} - {x}_c(t)+ \frac{i}{m\omega} \hat{p} \right), \\
			\hat{\tilde{a}}^\dagger(t) & = &\sqrt{\frac{m\omega}{2\hbar}} \left( \hat{x} - {x}_c(t)- \frac{i}{m\omega} \hat{p} \right).
		\end{eqnarray}
	\end{subequations}
	The Lindblad dissipator takes the form (see Appendix \ref{sec:translation} for details of the derivation):
	\begin{eqnarray}
		\label{eq:Lindblad-dynamics}
		{\mathcal{D}}_L
		& = &\frac{\gamma}{2}\bar n\left(  \hat{\tilde{a}}^\dagger (t)\hat{\rho}(t) \hat{\tilde{a}}(t) - \frac{1}{2}\left\{\hat{\tilde{a}}(t)\hat{\tilde{a}}^\dagger(t) ,\hat{\rho} (t)\right\} \right)  \\
		& + &\frac{\gamma}{2}(\bar n + 1) \left(  \hat{\tilde{a}} (t)\hat{\rho}(t) \hat{\tilde{a}}^\dagger(t)- \frac{1}{2} \left\{\hat{\tilde{a}}^\dagger (t)\hat{\tilde{a}}(t) ,\hat{\rho}(t) \right\} \right) \nonumber,
	\end{eqnarray}
	where $\left\{\hat{A}, \hat{B} \right\} = \hat{A}\hat{B}+\hat{B}\hat{A}$ for any two operators $\hat A$ and $\hat B$ acting in the Hilbert space of the quantum particle. We note that for the mathematical consistency of the suggested dissipator, $\dot{x}_c(t)$ must be a continuous function so that the driving protocol is continuous and differentiable in time (the necessity of this condition is discussed in Appendix~\ref{sec:activation}). The process $x_c(t)$ set by Eq.~\eqref{eq:AOUP} satisfies this condition.
	
	The dissipator \eqref{eq:Lindblad-dynamics} describes the dynamics emerging from the thermal contact of an oscillator with a thermal bath at temperature $T$. Here, 
	$\bar n$ is the expectation value of the oscillator number operator at thermal equilibrium, 
	\begin{equation}
		\bar n = \frac{1}{\exp(\hbar\omega/k_B T)-1}.
	\end{equation}
	where $k_B$ is the Boltzmann constant, and $\gamma$ is the thermalization rate \cite{Englert_2002}.
	
	In this work, we will analyze the interplay between the classical colored noise induced by the fluctuations of the potential and thermalization for thermalization rates.
	In what follows, we rewrite the dissipation strength using the gain and loss rates $\nu_{\pm}$, defined as 
	
	\begin{subequations}
		\begin{equation}
			\nu_{\pm} = \frac{\gamma}{2}\left(\bar n + \frac{1}{2}\right) \mp \frac{\gamma}{2}.
		\end{equation}
		with thermalization rate then regulated as
		\begin{equation}
			\gamma = 2(\nu_--\nu_+),
			\label{eq:diss}
		\end{equation}
		and temperature as
		\begin{equation}
			T = \frac{\hbar\omega}{k_B \ln\left(\frac{\nu_-}{\nu_+} \right)}.
			\label{eq:temp}
		\end{equation}
	\end{subequations}
	
	The dissipator \eqref{eq:Lindblad-dynamics} can be conveniently rewritten as
	\begin{eqnarray}
		{\mathcal{D}}_{L}\!& = &\!\nu_+\!\left(  \hat{\tilde{a}}^\dagger(t) \hat{\rho}(t) \hat{\tilde{a}}(t) - \frac{1}{2}\left\{\hat{\tilde{a}}(t)\hat{\tilde{a}}^\dagger(t) ,\hat{\rho}(t) \right\} \right) \nonumber \\
		& + &\!\nu_-\!\left(  \hat{\tilde{a}}(t) \hat{\rho}(t) \hat{\tilde{a}}^\dagger(t) - \frac{1}{2} \left\{\hat{\tilde{a}}^\dagger(t) \hat{\tilde{a}}(t) ,\hat{\rho}(t) \right\} \right)\!.
	\end{eqnarray}
	
	A convenient way to visualize the dynamics that the dissipator \eqref{eq:Lindblad-dynamics} generates makes use of the Wigner quasi-probability distribution $W(x,p)$, which provides a convenient phase-space representation of the state of a quantum particle \cite{Wigner:1932,Englert_2002,teVrugt2025}. 
	
\begin{figure}[ht!]
		\includegraphics[width=0.9\columnwidth]{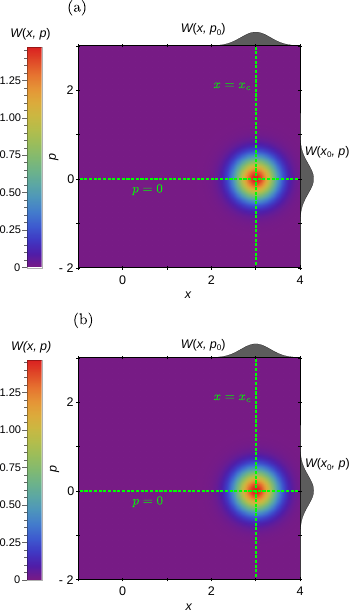}
		\caption{Steady-state Wigner function $W(x,p)$ for Lindblad master equation \eqref{eq:steady}. The position of the trapping potential is fixed at $x_c = 3$ for (a) weak dissipation and (b) strong dissipation. The amplitude of the Wigner function values is given by the corresponding color representation; the graphs on the left and right correspond to the peak of the Wigner function at the extremum in $x$ and $p$, respectively. The green dashed lines indicate the position ($x=x_c$) of the fixed ($p=0$) harmonic potential. The thermalization rate $\gamma = 2(\nu_--\nu_+)$ \eqref{eq:diss} is varied at fixed temperature $ T = \hbar\omega/\left(k_B \ln\left[\nu_-/\nu_+\right]\right)$ \eqref{eq:temp} with $\nu_-/\nu_+ = 100$ for $\nu_+ = 10^{-4}$ (weak) and $\nu_+=10^{-1}$ (strong).
		}
		\label{fig:wigner_dynamic}
	\end{figure}

	In the Wigner representation, the quantum master equation~\eqref{eq:Lindblad-dynamics} corresponds to a Fokker-Planck equation for $W(x,p)$, providing an alternative but fully equivalent description of the same quantum dynamics (see Appendix~\ref{sec:lin} for the derivation). The Fokker-Planck equation then reads:
	\begin{subequations}
		\label{eq:tFPE}
		\begin{eqnarray}
			\partial_t W \!& = &\!\sum_{i=x,p}\partial_i(f_i W) + \sum_{i,j=x,p}\partial^2_{ij}(g_{ij}W), \\
			f_x\!& = &\!-\frac{p}{m}\!+\!\frac{\gamma}{4}(x-x_c),\, f_p\!=\!m\omega^2(x-x_c)\!+\!\frac{\gamma}{4}p\\
			g_{xx}\! & = &\!\frac{\hbar\gamma}{8m\omega}\coth\!\left(\!\frac{\hbar \omega}{2k_B T}\!\right),\;g_{pp}\! =\!\frac{\hbar\gamma m\omega}{8}\coth\!\left(\!\frac{\hbar \omega}{2k_B T} \!\right),\nonumber \\
			\;
		\end{eqnarray}
	\end{subequations}
	where $g_{xp}=g_{px}=0$.
	We note that the Wigner Fokker-Planck equation contains terms that are not present in a dissipative classical harmonic oscillator (CHO),
	\begin{eqnarray}
		m\ddot{x}_{\rm cho}(t) &+ &\gamma\dot{x}_{\rm cho}(t) \\
		&=& -m\omega^2(x_{\rm cho}(t)-x_c) + \sqrt{\gamma k_B T}\eta(t)\,,\nonumber
	\end{eqnarray}
	
	for which the Fokker-Planck equation reads \cite{Tanimura/Wolynes:1991}:
	\begin{subequations}
		\label{eq:cFPE}
		\begin{eqnarray}
			\partial_t P_{\rm cho} \!& = &\!\sum_{i=x,p}\partial_i(f_i P_{\rm cho}) + \!\sum_{i,j=x,p}\partial^2_{ij}(g_{ij}P_{\rm cho}), \\
			f_x \!& = & \! -\frac{p}{m}, f_p = m\omega^2(x-x_c) + \gamma p,\,\\
			g_{pp}\! & = &\!\gamma k_B T,\, g_{xx} = g_{xp} = g_{px} = 0.\;
		\end{eqnarray}
	\end{subequations}
	To make the extremum points of the steady-state probability distribution explicit, let us for the moment assume that $x_c$ is constant so that the probability distribution of the quantum particle's positions is centered at $x_c$ in the steady state. 
	For fixed $x_c$, Eq.\ \eqref{eq:tFPE} has the stationary solution
	\begin{eqnarray}
		\label{eq:steady}
		&&W(x,p)\!\nonumber\\
		& = &\mathcal{N}\exp\left[-\tanh\left(\!\frac{\hbar \omega}{2 k_B T} \!\right)\!\left( \frac{m\omega (x\!-\!x_c)^2}{\hbar}\!+\!\frac{p^2}{m\hbar\omega}\right) \right ]\,.\nonumber\\
		\;
	\end{eqnarray}
	The steady-state Wigner function, shown in Fig.~\ref{fig:wigner_dynamic} in this case has the extremum of the distribution at $x=x_c$ and $p=0$ for both weak (Fig.~\ref{fig:wigner_dynamic}(a)) and strong (Fig.~\ref{fig:wigner_dynamic}(b)) dissipation strengths. Note that in our numerical studies $x_c=x_c(t)$ evolves with time; consequently, the particle is constantly driven away from the instantaneous trap center and is exposed to the colored noise mediated by the trap's motion.
	
	Although the Lindblad master equation with dissipator \eqref{eq:Lindblad-dynamics} correctly reproduces active motion even under strong dissipation, it does not display classical active motion in the classical limit. In fact, the structure of \eqref{eq:tFPE} is different from that of \eqref{eq:cFPE} in the limit $\hbar \to 0$, specifically in the force fields $f_i$. 
	
	\subsubsection{Agarwal dissipator}
	
	As an alternative to Lindblad dynamics, we consider what we denote by ``Agarwal dissipator'' \cite{Agarwal:1970, Agarwal:1971, Agarwal/Chaturverdi:2013, Lee/etal:2020}:
	\begin{eqnarray}
		\label{eq:Agarwal}
		{\mathcal{D}}_A & \equiv & {\mathcal{D}}(\hat{\rho}(t)) \\
		& = & - \frac{i \gamma}{8\hbar} [\hat{x}, \{ \hat{p}, \hat{\rho}(t) \}] 
		- \frac{\gamma m \omega}{4\hbar} \left( \bar{n} + \frac{1}{2} \right) [\hat{x}, [\hat{x}, \hat{\rho}(t)]].\nonumber
	\end{eqnarray}
	This dissipator recovers the well-known Caldeira-Leggett model in the high-temperature limit \cite{Caldeira/Legget:1983}. We note that dissipator \eqref{eq:Agarwal} is translationally invariant (see Appendix\ \ref{sec:translation}).
	
	The corresponding Wigner function representation reads (see Appendix~\ref{sec:Agarwal} for the derivation):
	\begin{subequations}
		\label{eq:FPE-abarwal}
		\begin{eqnarray}
			\partial_t P \!& = &\!\sum_{i=x,p}\partial_i(f_i P) + \sum_{i,j=x,p}\partial^2_{ij}(g_{ij}P), \\
			f_x \!& = & \! -\frac{p}{m}, f_p = m\omega^2(x-x_c) + \gamma p\,\\
			g_{pp}\! & = &\!\frac{\hbar\gamma m\omega}{8}\coth\left(\frac{\hbar \omega}{2k_B T}\right), g_{xx} = g_{xp} = g_{px} = 0.\nonumber\\
			\;
		\end{eqnarray}
	\end{subequations}
	which is analogous to the classical dissipative harmonic oscillator \eqref{eq:cFPE},\ with both the force fields $f_i$ and diffusion coefficient $g_{ij}$ identical between the classical and quantum cases in the limit $\hbar \to 0$.
	
	The stationary solution of Eq.~\eqref{eq:FPE-abarwal} for fixed $x_c$ is given by Eq.\ \eqref{eq:steady} and therefore identical to that of the Lindblad master equation. However, while both Fokker-Planck equations have the same steady-state for $x_c$ constant, their dynamics differ. This leads to further discrepancies under a time-dependent driving protocol $x_c(t)$. In fact, in the dynamics governed by the Lindblad master equation with dissipator \eqref{eq:Lindblad-dynamics}, both the Hamiltonian part and the dissipator ${\mathcal{D}}$ drive the quantum particle toward the minimum of the moving trap $x_c(t)$. In contrast, for the Agarwal dissipator, the dissipator ${\mathcal{D}}$ acts purely as a source of friction, as is evident from the corresponding Wigner function representations \eqref{eq:tFPE}, \eqref{eq:FPE-abarwal}.
	
	Thus, even though the dissipator \eqref{eq:Agarwal} does not satisfy the Lindblad theorem and may violate positiveness of the density matrix for some of the stochastic trajectories (see Appendix~\ref{sec:simulation} for an example of such a violation), it is designed to yield the correct thermodynamics. This makes it well-suited for studying classical systems approaching thermodynamic equilibrium where quantum effects become significant.

	\subsection{Numerical procedure}
	
	In order to assess quantum active behavior in our model, we compare the numerical results for the quantum mean squared displacements (MSDs), defined as \cite{Antonov/etal:2025}
	
	\begin{equation}
		\MSD(t) = \overline{\Tr(\hat{\rho}(t)\hat{x}^2)} - \Tr(\hat{\rho}(0)\hat{x}^2),
		\label{eq:quantum-MSD}
	\end{equation}
	where $\overline{\cdots}$ denotes the average over prescribed trajectories $x_c(t)$. In \eqref{eq:quantum-MSD}, we first evaluate the trace of the observable (i.e.,\ perform the quantum averaging for given instantaneous $x_c(t)$), and only afterwards average over the prescribed trajectories $x_c(t)$. 
	This ordering is natural since $x_c(t)$ is a classical variable, following a classical probability distribution. Therefore, Eq.~\eqref{eq:varrho} involves purely classical sampling, 
	and the order of averaging in Eq.\ \eqref{eq:quantum-MSD} is irrelevant. 
	
	For numerical simulations, we use the predictor-corrector scheme (see Appendix~\ref{sec:simulation} for details; the full implementation is available at \footnote{The Python implementation is available from GitHub repository,~\href{https://github.com/apantonov/quantum-active}{https://github.com/apantonov/quantum-active}.}) 
	and the averaging is performed over classical trajectories. The quantum particle is initially taken to be in the ground state $\hat{\rho} = |0\rangle\langle0|$, where $\ket{0}$ denotes the normalized ground state of the quantum harmonic oscillator, corresponding to the lowest-energy eigenstate of the Hamiltonian. The stochastic trajectories $x_c$ are initialized at $x_c(0) = 0$ and $\dot{x}_c(0)=0$.
	
	\subsection{Potential experimental realization}
	A possible realization of the model considered here is a quantum-mechanical particle in a trap centered at position $x_c(t)$. The trajectory of the center~$\{x_c(t)\}$ can be generated by a computer. 
	By repeating this procedure for a set of $\{x_c(t)\}$, the average value of the quantum mean squared displacement, explained in the following subsection, can be measured. A sketch of such experimental realization is shown in Fig.~\ref{fig:setup}. We note that equations similar to the dissipator Eq.\ \eqref{eq:Lindblad-dynamics} are used to describe the laser cooling dynamics of a particle trapped by a harmonic oscillator, such as for instance a single trapped ion \cite{Eschner:2003}. Here, the thermalization rate and the temperature can be varied by changing the intensity and detuning of the laser \cite{Stenholm:1986,Eschner:2003,Morigi:2003}.
	
	\section{Results for the mean-squared displacements}
	\label{sec:results}
	
\begin{figure}[hb!]
		\includegraphics[width=0.9\columnwidth]{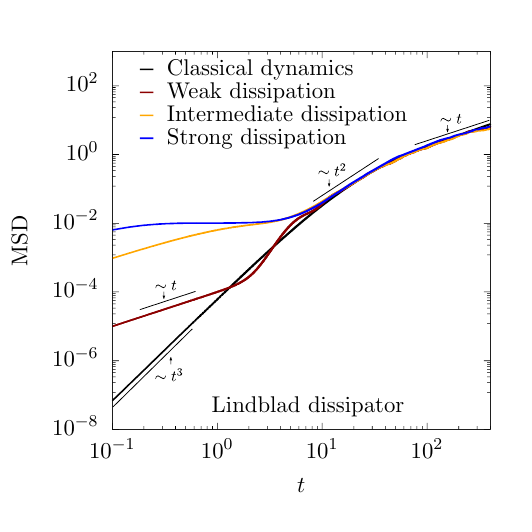}
		\caption{MSD of a quantum-active particle for Lindblad dissipator \eqref{eq:Lindblad-dynamics} for weak, intermediate and strong dissipation respectively. The parameters of $x_c$ are $D = 0.01, \tau = 10$, and the corresponding classical $\MSD = \overline{x_c^2(t)}$ of these trajectories is shown as a black line. The thermalization rate $\gamma = 2(\nu_--\nu_+)$ \eqref{eq:diss} is varied at fixed temperature $ T = \hbar\omega/\left(k_B \ln\left[\nu_-/\nu_+\right]\right)$ \eqref{eq:temp} with $\nu_-/\nu_+ = 100$ for $\nu_+ = 10^{-4}$ (weak), $\nu_+ = 10^{-2}$ (intermediate) and $\nu_+=10^{-1}$ (strong).}
		\label{Fig:5}
	\end{figure}

	In this section, we present the time evolution of the MSD, Eq.\ \eqref{eq:quantum-MSD} by comparing the predictions of both forms of dissipators considered in this paper. We analyze different thermalization rates $\gamma$, which we refer to as dissipation strengths by analogy with classical mechanics, since $\gamma$ plays the role of dissipation in the corresponding classical system \eqref{eq:cFPE}. We choose $\sqrt{\hbar / m\omega}$ and $\omega^{-1}$ as units of length and time, respectively. In Fig.~\ref{Fig:5}, we first compare the MSD of Eq.\ \eqref{eq:quantum-MSD} of the master equation with Lindblad dissipator \eqref{eq:Lindblad-dynamics}, using the classical MSD of the driving protocol, $\overline{x_c^2(t)}$, as a reference. We recall that $x_c(t)$ in Eq.~\eqref{eq:AOUP} corresponds to a \textit{passive} inertial particle, exhibiting $t^3$-scaling at short times before crossing over to diffusion \cite{Huy/etal:2021}. When
	$x_c(t)$ acts as a colored-noise input to the moving potential, it drives the particle in a way that breaks the detailed balance, allowing us to study how quantum dissipators influence motion that mimics \textit{active} dynamics.
	
	For weak thermalization rates, $\gamma\ll 1$ the MSD initially shows diffusive behavior, followed by a crossover to parabolic growth and later back to linear diffusion, thereby exhibiting properties of an active particle \cite{Bechinger/etal:2016} and consistent with the predictions for low temperatures and weak thermalization \cite{Antonov/etal:2025}. However, as $\gamma$ increases, the effect of the dissipator ${\mathcal{D}}$ becomes more significant for $\gamma \gg 1$, so that the initial diffusive regime is no longer observable. Nevertheless, the ballistic-to-linear scaling of the MSD for $t \gtrsim \tau$ remains evident across the entire range of dissipation rates.

	\begin{figure}[hb!]
		\includegraphics[width=0.9\columnwidth]{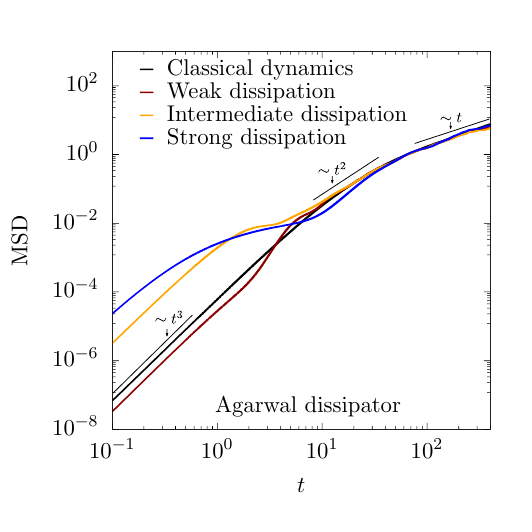}
		\caption{MSD of a quantum-active particle with Agarwal dissipator \eqref{eq:Agarwal} for weak, intermediate and strong dissipation respectively. The parameters of $x_c$ are $D = 0.01, \tau = 10$, and the corresponding classical $\MSD = \overline{x_c^2(t)}$ of these trajectories is shown as a black line. The thermalization rate $\gamma = 2(\nu_--\nu_+)$ \eqref{eq:diss} is varied at fixed temperature $ T = \hbar\omega/\left(k_B \ln\left[\nu_-/\nu_+\right]\right)$ \eqref{eq:temp} with $\nu_-/\nu_+ = 100$ for $\nu_+ = 10^{-4}$ (weak), $\nu_+ = 10^{-2}$ (intermediate) and $\nu_+=10^{-1}$ (strong).
		}
		\label{Fig:6}
	\end{figure}
	
\begin{figure}[hb!]
		\includegraphics[width=0.9\columnwidth]{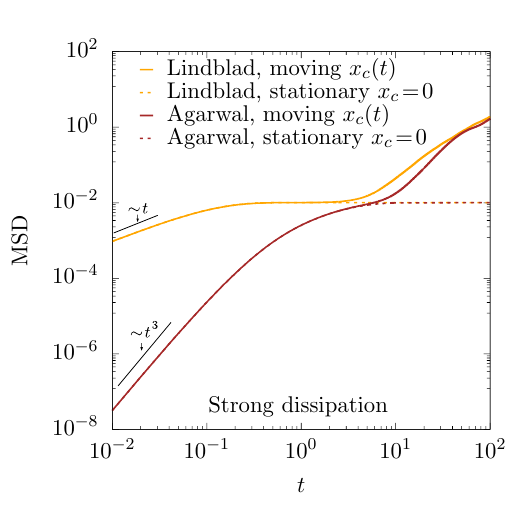}
		\caption{MSD of a quantum-active particle with Agarwal \eqref{eq:Agarwal} (red) and Lindblad (orange) dissipators \eqref{eq:Lindblad-dynamics} for strong dissipation $\nu_+=10^{-1},\, \nu_-=10^1$. The stationary case, $x_c = 0$ (dashed lines) is compared to the dynamical one given by Eq.~\eqref{eq:AOUP} with $D = 0.01, \tau = 10$, revealing identical MSD evolution for $t\ll \tau$. The thermalization rate is $\gamma = 2(\nu_--\nu_+)$ \eqref{eq:diss}, and the temperature is $ T = \hbar\omega/\left(k_B \ln\left[\nu_-/\nu_+\right]\right)$ \eqref{eq:temp}.
		}
		\label{Fig:7}
	\end{figure}

	The differences between the predictions for the Lindblad and of the Agarwal dissipators are visible by comparing Fig.~\ref{Fig:5} and Fig.~\ref{Fig:6} and become evident at the dissipation time scale of the classical particle $\tau$. At long time scales, $t \gtrsim \tau$, similar to the Lindblad dissipator, the dynamics of quantum particles also follows that of the center of the trapping potential and exhibits the features of the active motion across all dissipation rates.
	However, the different nature of the considered dissipators leads to distinct MSD regimes at short time scales, $t \ll \tau$. In the Lindblad case, quantum fluctuations induced by the dissipator -- generated by the creation and annihilation operators of the Lindblad dissipator -- produce a diffusive regime (Fig.~\ref{Fig:5}). By contrast, this short-time diffusive regime is absent for the Agarwal dissipator (Fig.~\ref{Fig:6}): the quantum particle simply follows the stochastic trajectory $x_c(t)$ with an inertial delay and has the same $t^3$-scaling as the driving protocol. In Fig.~\ref{Fig:7}, we demonstrate that these distinctions do not depend on the nature of the driving protocol $x_c(t)$ and are due to initial wavepacket broadening arising from the dissipator. The demonstrated results are robust to moderate changes of $D$ and $\tau$ in the active protocol. A detailed analysis is presented in \cite{Sangyun/etal:2026}.
	
	\section{Conclusions and outlook}
	\label{sec:conclusion}
	In this work, we studied how different types of quantum dissipators influence the emergence of active-like dynamics of a quantum particle, aiming to understand how the environment affects the characteristics of motion reminiscent of classical active matter. We have compared the dynamics predicted by two forms of time-local master equations and explored the interplay between dissipation strength, external driving, and quantum effects on determining the mean squared deviation. The dissipators individually satisfy different requirements: either positivity or the correct thermodynamic properties.
	We find that the quantum particle exhibits typical behaviors of active particles, namely ballistic motion at intermediate times ($\textrm{MSD} \propto t^2$) and active diffusion at late times ($\textrm{MSD} \propto t$). Importantly, after averaging over many realizations of the actuating potential, the system is isotropic, such that the external field does not break the symmetry of the system, which is typical for active matter systems \cite{Vrugt/etal:2025}. Beyond that, the short-time dynamics is governed by quantum effects and depends, in particular, on the chosen form of the dissipator.
	
	Our findings could be experimentally tested in
	cold-atom systems, in particular in moving optical traps with controlled fluctuations of the laser intensity \cite{Savard/etal:1997}. From a theoretical perspective, developing strategies for controlling quantum particles in stochastic \cite{Beyer/Paul:2021} and dissipative \cite{Englert_2002} environments will be crucial for bridging the gap between abstract models and experimentally accessible setups \cite{Burgardt/etal:2026}.
	
	\section*{Acknowledgments}
	M.t.V.\ is funded by the Deutsche Forschungsgemeinschaft (DFG, German Research Foundation) -- SFB 1551, Project-ID  464588647. G.M. acknowledges funding from the QuantERA II Programme (project ``QNet: Quantum transport, metastability, and neuromorphic applications in Quantum Networks"), which has received funding from the EU's Horizon 2020 research and innovation programme under Grant Agreement No. 101017733, as well as from the Deutsche Forschungsgemeinschaft DFG (Project ID 532771420). The authors acknowledge discussions with L.\ Chomaz, T.\ Pfau, P.\ Schmelcher, and A.\ Widera. S.L. thanks R. Kosloff for a discussion on modeling an open quantum system.

	\newpage
	\appendix
	
	\begin{widetext}
		\section{Wigner-Weyl transform}
		\label{sec:Weyl}
		
		Here, we show how to compute the Wigner-Weyl transforms used in the main text of this manuscript. For a pedagogical introduction to the Wigner-Weyl transform, we refer the reader to Refs.~\cite{zachos2005quantum,teVrugt2025}.
		\subsection{Lindblad dissipator}
		\label{sec:lin}
		
		We consider here the creation-annihilation operators in the form
		\begin{subequations}
			\label{eq:CA1}
			\begin{eqnarray}
				\hat{\tilde{a}}(t) & = & \sqrt{\frac{m\omega}{2\hbar}} \left( \hat{x} - \delta\hat{x}_c(t)+ \frac{i}{m\omega}\hat{p} \right), \\
				\hat{\tilde{a}}^\dagger(t) & = &\sqrt{\frac{m\omega}{2\hbar}} \left( \hat{x} - \delta\hat{x}_c(t)- \frac{i}{m\omega} \hat{p} \right),
			\end{eqnarray}
		\end{subequations}
		Here, we introduced a parameter $\delta$ in order to cover both the scenario studied in the main text ($\delta=1$, cf. Eq.~\eqref{eq:creation-annihilation}) and the scenario studied in Ref.\ \cite{Antonov/etal:2025} ($\delta=0$), see~Appendix \ref{sec:form} for a discussion how these forms are related.
		
		For the Hamiltonian part, the Wigner-Weyl transform is well known \cite{Hillery/etal:1984}. Here we first show the full expression with the Moyal product; since we are dealing only with harmonic oscillators, higher-order corrections vanish and can be neglected.
		
		\begin{subequations}
			\label{eq:Wigner-Hamiltonian}
			\begin{eqnarray}
				\left(\partial_t \rho\right)_W&  \equiv &\partial_tW(x,p,t),\\
				-\frac{i}{\hbar}\left(\left[\hat{H}(t),\rho\right]\right)_W & = & H \star W - W \star H \nonumber = \partial_xH\partial_pW-\partial_pH\partial_xW+O(\hbar^2),
			\end{eqnarray}   
		\end{subequations}
		where $\star$ is the Moyal product defined as
		
		\begin{equation}
			A \star B \equiv A_W\, \exp\Bigg[\frac{i\hbar}{2}\Big(\overleftarrow{\partial_x}\,\overrightarrow{\partial_p} - \overleftarrow{\partial_p}\,\overrightarrow{\partial_x}\Big)\Bigg] B_W,
		\end{equation}
		or equivalently, expanded in powers of $\hbar$:
		\begin{equation}
			A \star B = A_W B_W + \frac{i\hbar}{2} \{A_W, B_W\} - \frac{\hbar^2}{8} (\partial_x^2 A_W \, \partial_p^2 B_W - 2 \partial_{xp}^2 A_W \, \partial_{xp}^2 B_W + \partial_p^2 A_W \, \partial_x^2 B_W) + \dots,
		\end{equation}
		where $\{f,g\}=\partial_xf\partial_pg - \partial_px\partial_xg$ are the Poisson brackets related to the Moyal product as:
		
		\begin{equation}
			{\frac {1}{i\hbar }}(f\star g-g\star f)=\{f,g\}+O(\hbar ^{2}).
		\end{equation}
		Here and in Eq.~\eqref{eq:Wigner-Hamiltonian}, all corrections of order $O(\hbar^2)$ are equal to zero for the harmonic potential considered.
		
		For the Lindblad dissipator,
		\begin{eqnarray}
			\left(\hat{A}\hat{\rho}\hat{B}\right)_W & = & (A_W \star W)\star B_W\\
			& = &A_W W B_W + \frac{i\hbar}{2}\left[\{A_W W, B_W\} + \{A_W, W\}B_W \right] - \frac{\hbar^2}{4}\left\{\{A_W, W\}, B_W \right\} + O(\hbar^3).
			\label{eq:triple-product}
		\end{eqnarray}
		
		Given that operators $\hat{A},\hat{B}$ in the dissipator ${\mathcal{D}}$ are linear in $x,p$, Eq.~\eqref{eq:triple-product} reduces to:
		
		\begin{eqnarray}
			\left(\hat{A}\hat{\rho}\hat{B}\right)_W &  = & A_W W B_W\nonumber\\
			& + & \frac{i\hbar}{2}\left((\partial_xA_W\partial_pB_W-\partial_pA_W\partial_xB_W)W + A_W(\partial_xW\partial_pB_W - \partial_pW\partial_xB_W) + B_W(\partial_xA_W\partial_pW - \partial_pA_W\partial_xW) \right)\nonumber\\
			& - & \frac{\hbar^2}{4}\left(\partial^2_{xp}W(\partial_x A_W\partial_p B_W+ \partial_p A_W\partial_x B_W) - \partial_p B_W\partial_p A_W \partial^2_xW-\partial_xA_W\partial_xB_W\partial^2_pW\right).
			\label{eq:wigner-2}
		\end{eqnarray}
		
		Thus, the triple products relevant in our case are
		
		\begin{eqnarray}
			\left(\hat{\tilde a}^\dagger\hat{\rho}\hat{\tilde a}\right)_W &  = & |\tilde a|^2W\nonumber\\
			& + & \frac{i\hbar}{2}\left((\alpha\beta-(-\beta)\alpha)W + \tilde a^\dagger(\partial_xW\beta - \partial_pW\alpha) + \tilde a(\alpha \partial_pW - (-\beta)\partial_xW) \right)\nonumber\\
			& - & \frac{\hbar^2}{4}\left(\partial^2_{xp}W(\alpha\beta - \beta \alpha) + \beta^2 \partial^2_xW-\alpha^2\partial^2_pW\right)=\nonumber\\
			& = & |\tilde a|^2W - \frac{1}{2}\left(W + p\partial_pW + (x-\delta x_c)\partial_xW\right) +\frac{m\omega\hbar}{8}\partial_x^2W + \frac{\hbar}{8m\omega}\partial_p^2W;\\
			\left(\hat{\tilde a}\hat{\rho}\hat{\tilde a}^\dagger\right)_W &  = & |\tilde a|^2W\nonumber\\
			& + & \frac{i\hbar}{2}\left((-\alpha\beta-\beta\alpha)W + \tilde a(-\partial_xW\beta - \partial_pW\alpha) + \tilde a^{\dagger}(\alpha \partial_pW - \beta\partial_xW) \right)\nonumber\\
			& - & \frac{\hbar^2}{4}\left(\partial^2_{xp}W(-\alpha\beta + \beta \alpha) + \beta^2 \partial^2_xW-\alpha^2B\partial^2_pW\right)=\nonumber\\
			& = & |\tilde a|^2W + \frac{1}{2}\left(W + p\partial_pW + (x-\delta x_c)\partial_xW\right) +\frac{m\omega\hbar}{8}\partial_x^2W + \frac{\hbar}{8m\omega}\partial_p^2W;
		\end{eqnarray}
		where we introduced $\alpha = \sqrt{m\omega/2\hbar},\, \beta = i/\sqrt{2m\omega \hbar}$. 
		
		For the anti-commutators, we obtain
		
		\begin{eqnarray}
			\frac{1}{2}\left\{\hat{\tilde a} \hat{\tilde a}^\dagger ,\hat{\rho} \right\} & = &\left(|\tilde a|^2 + \frac{i\hbar}{2}(\partial_x \tilde a\partial_p \tilde a^\dagger-\partial_p \tilde a\partial_x \tilde a^\dagger) \right) W - \frac{\hbar^2}{8}(\partial_x^2|\tilde a|^2\partial_p^2W-\partial_p^2|\tilde a|^2\partial^2_xW)\nonumber\\
			&=& (|\tilde a|^2 + \frac{1}{2})W - \frac{\hbar^2}{4}(\alpha^2 \partial_p^2W-\beta^2\partial^2_xW),\\
			\frac{1}{2}\left\{\hat{\tilde a}^\dagger \hat{\tilde a} ,\hat{\rho} \right\} & = &\left(|\tilde a|^2 + \frac{i\hbar}{2}(\partial_x \tilde a^\dagger\partial_p \tilde a-\partial_p \tilde a^\dagger\partial_x \tilde a) \right) W - \frac{\hbar^2}{8}(\partial_x^2|\tilde a|^2\partial_p^2W-\partial_p^2|\tilde a|^2\partial^2_xW)\nonumber\\
			&=& (|\tilde a|^2 - \frac{1}{2})W - \frac{\hbar^2}{4}(\alpha^2 \partial_p^2W-\beta^2\partial^2_xW).
		\end{eqnarray}
		
		Finally, we arrive at
		
		\begin{eqnarray}
			({\mathcal{D}})_W & = & \nu_+\Bigg( - \frac{1}{2}\left(2W + p\partial_pW + (x-\delta x_c)\partial_xW\right) +\frac{m\omega\hbar}{4}\partial_x^2W + \frac{\hbar}{4m\omega}\partial_p^2W\Bigg)\nonumber\\
			& + & \nu_-\Bigg(\frac{1}{2}\left(2W + p\partial_pW + (x-\delta x_c)\partial_xW\right) +\frac{\hbar}{4m\omega}\partial_x^2W + \frac{m\omega\hbar}{4}\partial_p^2W\Bigg).
			\label{eq:Lind}
		\end{eqnarray}
		
		The resulting Fokker-Planck equation, obtained by combining \eqref{eq:Wigner-Hamiltonian} with \eqref{eq:Lind}, is given by
		
		\begin{subequations}
			\label{eq:FPE1}
			\begin{eqnarray}
				\partial_t W \!& = &\!\sum_{i=x,p}\partial_i(f_i W) + \sum_{i,j=x,p}\partial^2_{ij}(g_{ij}W), \\
				f_x & = & -\frac{p}{m} + \frac{\gamma}{4}(x - \delta x_c), \quad f_p= m\omega^2(x-x_c) + \frac{\gamma}{4}p\\
				g_{xx} & = &\frac{\hbar\gamma}{8m\omega}\coth\left(\frac{\hbar \omega}{2k_B T} \right),\quad g_{pp}=\frac{\hbar\gamma m\omega}{8}\coth\left(\frac{\hbar \omega}{2k_B T} \right),\quad g_{xp}=g_{px}=0.\nonumber\\
				\;
			\end{eqnarray}
		\end{subequations}

		Here, we have used the following relations:
		
		\begin{eqnarray}
			\nu_- & = & \frac{\gamma}{2}(\bar{n} + 1), \ \nu_+ = \frac{\gamma}{2}\bar{n}, \\
			\nu_- - \nu_+ & = & \frac{\gamma}{2}, \ \frac{\nu_-}{\nu_+} = 1 + \bar{n}^{-1} = \exp\left(\frac{\hbar \omega}{k_B T}\right),\label{eq:gamma}\\
			\nu_- + \nu_+ & = & \gamma\left(\bar{n} + \frac{1}{2}\right) = \frac{\gamma}{2}\coth\left(\frac{\hbar \omega}{2k_B T} \right).
			\label{eq:refer}
		\end{eqnarray}
		
		\subsection{Role of the Lindblad dissipator form}
		\label{sec:form}
		For the Lindblad dissipator, Ref.\ \cite{Antonov/etal:2025} has used instead of Eq.~\eqref{eq:tFPE} the alternative form:
		\begin{eqnarray}
			\label{eq:Lindblad-static}
			{\mathcal{D}}_{SL} = {\mathcal{D}}(\hat{\rho}(t))& = &\frac{\gamma}{2}\bar n\left(  \hat{a}^\dagger \hat{\rho}(t) \hat{a} - \frac{1}{2}\left\{\hat{a}\hat{a}^\dagger ,\hat{\rho}(t) \right\} \right)  \\
			& + &\frac{\gamma}{2}(\bar n + 1) \left(  \hat{a} \hat{\rho}(t) \hat{a}^\dagger - \frac{1}{2} \left\{\hat{a}^\dagger \hat{a} ,\hat{\rho}(t) \right\} \right),\nonumber
		\end{eqnarray}
		with the static ladder operators
		\begin{subequations}
			\label{eq:ladder-static}
			\begin{eqnarray}
				\hat{a} & = & \sqrt{\frac{m\omega}{2\hbar}}\left(\hat{x} + \frac{i}{m\omega}\hat{p}\right), \\
				\hat{a}^\dagger & = & \sqrt{\frac{m\omega}{2\hbar}}\left(\hat{x} - \frac{i}{m\omega}\hat{p}\right).
			\end{eqnarray}
		\end{subequations}
		This variant is appropriate for the case of small damping, which was the focus of interest in Ref.\ \cite{Antonov/etal:2025}. Here, we discuss why for the case of stronger damping one should use the form \eqref{eq:tFPE}, which can be derived from \eqref{eq:Lindblad-static} by applying a translation operator (see Appendix \ref{sec:translation}).
		
		The Fokker-Planck equation for the Wigner function corresponding to the dissipator \eqref{eq:Lindblad-static} reads
		\begin{subequations}
			\label{eq:fPE}
			\begin{eqnarray}
				\partial_t W \!& = &\!\sum_{i=x,p}\partial_i(f_i W) + \sum_{i,j=x,p}\partial^2_{ij}(g_{ij}W), \\
				f_x & = & -\frac{p}{m} + \frac{\gamma}{4}x, \quad f_p= m\omega^2(x-x_c) + \frac{\gamma}{4}p\\
				g_{xx} & = &\frac{\hbar\gamma}{8m\omega}\coth\left(\frac{\hbar \omega}{2k_B T} \right),\quad g_{pp}=\frac{\hbar\gamma m\omega}{8}\coth\left(\frac{\hbar \omega}{2k_B T} \right),\quad g_{xp}=g_{px}=0.\nonumber\\
				\;
			\end{eqnarray}
		\end{subequations}
		
		\begin{figure}[htp!]
			\vspace{-2ex}
			\centering\includegraphics[width=0.4\linewidth]{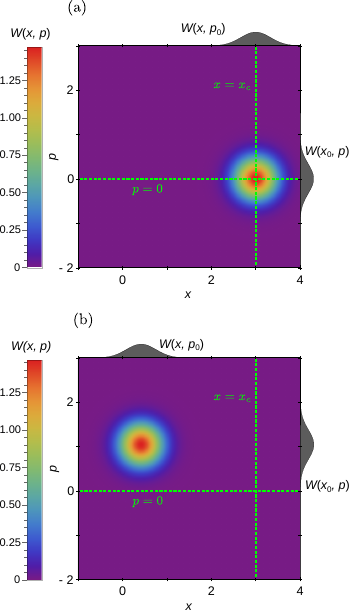}
			\caption{Steady-state Wigner function $W(x,p)$ for ``static'' Lindblad dissipator \eqref{eq:steady-L}. The position of the trapping potential is fixed at $x_c = 3$ for (a) weak dissipation and (b) strong dissipation. The amplitude of the Wigner function values is given by the corresponding color representation; the graphs on the left and right correspond to the peak of the Wigner function at the extremum in $x$ and $p$, respectively. The green dashed lines indicate the position ($x=x_c$) of the fixed ($p=0$) harmonic potential. The thermalization rate $\gamma = 2(\nu_--\nu_+)$ \eqref{eq:diss} is varied at fixed temperature $ T = \hbar\omega/\left(k_B \ln\left[\nu_-/\nu_+\right]\right)$ \eqref{eq:temp} with $\nu_-/\nu_+ = 100$ for $\nu_+ = 10^{-4}$ (weak) and $\nu_+=10^{-1}$ (strong).
			}
			\label{fig:wigner_static}
		\end{figure} 
		
		For fixed $x_c$, the steady-state of Eq.~\eqref{eq:fPE} reads:
		\begin{align}\label{eq:steady-L}
			W(x,p)
			= \mathcal N
			\exp{[
				-\frac{1}{2} ( q-  q_0)\cdot M\cdot ( q- q_0)
				]}
		\end{align}
		where
		\begin{align}
			M =& \begin{pmatrix}
				m\omega^2/\tilde T & 0\\
				0& 1/ m \tilde T
			\end{pmatrix}, \quad
			q= 
			\begin{pmatrix}
				x  \\
				p
			\end{pmatrix},
			\nonumber \\
			q_0=& \begin{pmatrix}
				x_0   \\
				p_0
			\end{pmatrix} =
			\begin{pmatrix}
				16\omega^2x_c/(\gamma^2+16\omega^2)   \\
				4\gamma m\omega^2 x_c/(\gamma^2+16\omega^2)
			\end{pmatrix}
			,\end{align}
		with $\tilde T \equiv \frac{\hbar \omega}{2}\coth{ \frac{\hbar \omega}{2k_BT} } $ and $\mathcal N = \tilde{T}/(2\pi\omega)$. 
		It is evident that the extremum of the position distribution $x = x_0$ depends on dissipation strength $\gamma$. For weak damping $\gamma \to 0$, it is at $x_0 \to x_c$, while for strong damping $\gamma \to \infty$, $x_0 \to 0$. For the steady-state Wigner function $W(x,p)$ shown in Fig.~\ref{fig:wigner_static} we indeed observe that the expected behavior near the shifted trap center $x = x_c$ is reproduced in the weak-damping limit (Fig.~\ref{fig:wigner_static}(a)), while driving the quantum particle close to the original position $x=0$ in the strong-damping limit (Fig.~\ref{fig:wigner_static}(b)). Moreover, except for the weak-damping limit $\gamma \to 0$, the steady-state momentum $p_0$ remains finite, which highlights the unphysical nature of the model with $\delta=0$. In this case, the ladder operators \eqref{eq:ladder-static} depend on position $x$ instead of the actual position of the harmonic oscillator $x-x_c$, which is correctly implemented by setting $\delta = 1$ in the main text (Eq.~\eqref{eq:creation-annihilation}). The inconsistency of the ``static'' dissipator $\mathcal{D}_{SL}$ \eqref{eq:Lindblad-static} is also evident in the mean-square displacements shown in Fig.~\ref{Fig:4}: for strong dissipation, the center of the
		particle distribution no longer follows the potential minimum and ceases to exhibit features of activity. By contrast, for the Lindblad dissipator $\mathcal{D}_L$ discussed in the main text ($\delta=1$, Eq.~\eqref{eq:Lindblad-dynamics}), the steady-state momentum vanishes, $p_0 = 0$, and the active-like behavior is preserved across all dissipation strengths (Fig.~\ref{Fig:5}).
		
		\begin{figure}[htp!]
			\vspace{-3ex}
			\includegraphics[width=0.4\linewidth]{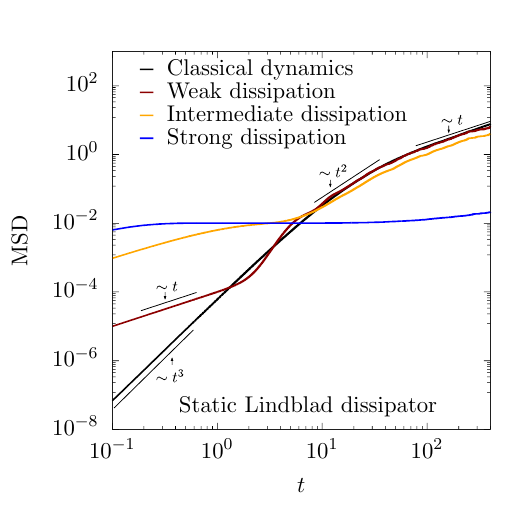}
			\vspace{-4ex}
			\caption{MSD of a quantum-active particle with ``static'' Lindblad dissipator Eq.~\eqref{eq:Lindblad-static} 
				for weak, intermediate and strong dissipation respectively. The parameters of $x_c$ are $D = 0.01, \tau = 10$, and the corresponding classical $\MSD = \overline{x_c^2(t)}$ of these trajectories is shown as a black line. The thermalization rate $\gamma = 2(\nu_--\nu_+)$ \eqref{eq:diss} is varied at fixed temperature $ T = \hbar\omega/\left(k_B \ln\left[\nu_-/\nu_+\right]\right)$ \eqref{eq:temp} with $\nu_-/\nu_+ = 100$ for $\nu_+ = 10^{-4}$ (weak), $\nu_+ = 10^{-2}$ (intermediate) and $\nu_+=10^{-1}$ (strong).}
			\label{Fig:4}
		\end{figure}
		
		\newpage
		\subsection{Agarwal dissipator}
		\label{sec:Agarwal}
		
		The Wigner-Weyl transform for the Agarwal dissipator is performed analogously by using manipulations performed in Eqs.~\eqref{eq:triple-product}-\eqref{eq:wigner-2}:
		
		\begin{eqnarray}
			\left(\hat{x} \hat{p} \hat{\rho}\right)_W & \!=\!&\!xpW + \frac{i\hbar}{2}\left(W + p\partial_p W - x\partial_x W\right);\\
			\left(\hat{x} \hat{\rho} \hat{p}\right)_W & \!=\!&\!xpW + \frac{i\hbar}{2}\left(W + p\partial_p W + x\partial_x W\right) - \frac{\hbar^2}{4}\partial_{xp}^2W;\nonumber \\
			\\
			\left(\hat{p} \hat{\rho} \hat{x}\right)_W& \!=\!&\!xpW - \frac{i\hbar}{2}\left(W + p\partial_p W + x\partial_x W\right) - \frac{\hbar^2}{4}\partial_{xp}^2W ;\nonumber \\
			\\
			\left(\hat{\rho} \hat{p} \hat{x}\right)_W & \!=\!&\!xpW + \frac{i\hbar}{2}\left(W - p\partial_p W + x\partial_x W\right);\\ 
			\left(\hat{x} \hat{x} \hat{\rho}\right)_W & \!=\!&\!x^2W + i\hbar\left(x \partial_p W\right);\\
			\left(\hat{\rho} \hat{x} \hat{x} \right)_W & \!=\!&\!x^2W - i\hbar\left(x \partial_p W\right);\\
			\left(\hat{x}\hat{\rho}  \hat{x} \right)_W & \!=\!&\!x^2W +\frac{\hbar^2}{4}\partial_p^2 W.
		\end{eqnarray}
		
		\begin{eqnarray}
			\label{eq:FPE-agarwal}
			{\mathcal{D}}_A = &-&\frac{i \gamma}{8\hbar} (\hat{x} \hat{p} \hat{\rho} + \hat{x} \hat{\rho} \hat{p} - \hat{p} \hat{\rho} \hat{x} - \hat{\rho} \hat{p}\hat{x})\nonumber\\
			&-&\frac{\gamma m \omega}{4\hbar} \left(\bar n + \frac{1}{2}\right) (\hat{x} \hat{x} \hat{\rho} - 2 \hat{x} \hat{\rho} \hat{x} + \hat{\rho} \hat{x} \hat{x})\nonumber\\
			& = &\frac{\hbar \gamma}{4} (W + p\partial_p W) + \frac{\hbar \gamma m \omega}{8}  \coth\left(\frac{\hbar\omega}{2k_B T}\right)\partial_p^2W.
		\end{eqnarray}
		
		Combining \eqref{eq:Wigner-Hamiltonian} with \eqref{eq:FPE-agarwal} yields \eqref{eq:FPE-abarwal}.
		
		\section{Translation of the quantum dissipators}
		\label{sec:translation}
		Here, we show how one can obtain the general Lindblad dissipator \eqref{eq:tFPE} by applying the translation generator $\hat T(x_c)$ to \eqref{eq:Lindblad-static}. Physically, this corresponds to taking into account the trap motion when modeling the bath dynamics. This is (also for readers from classical active soft matter physics) worth discussing here explicitly because it marks a contrast to the case of classical baths, which tend to have an Agarwal-like structure and thus do not require to take translations into account for bath models in this way.
		
		We consider the evolution equation \eqref{eq:evolution} with the static Lindblad dissipator \eqref{eq:Lindblad-static} translated by $x_c$, i.e., we apply the translation generator $\hat T(x_c)$, 
		
		\begin{equation}
			\label{eq:transl}
			\hat T(x_c)= e^{-ix_c \hat p /\hbar}
		\end{equation}
		to the Hamiltonian~
		\begin{align}
			\hat H_0 \equiv \frac{\hat p^2}{2m} + \frac{1}{2}m\omega^2 \hat x ^2, 
		\end{align}
		and the dissipator ${\mathcal{D}_{SL}}$. In other words, a state vector is redefined ($|\psi\rangle=\hat T |\psi\rangle $ and $\hat \rho = \hat T \hat \rho \hat T^\dagger $), and all operators are transformed as $\hat T\hat b \hat T^\dagger$.
		
		For the Lindblad dissipator \eqref{eq:Lindblad-static},
		\begin{eqnarray}
			{\mathcal{D}}_{SL}& = &\nu_- [ \hat a \hat \rho(t) \hat a^{\dagger} - \frac{1}{2} \left\{\hat a^{\dagger} \hat a,    \hat \rho(t)\right\} ]\nonumber \\
			& + &\nu_+  [ \hat a^{\dagger} \rho(t) \hat a - \frac{1}{2} \left\{\hat a \hat a^{\dagger},    \hat \rho(t)\right\}  ],
		\end{eqnarray}
		the creation/annihilation operators are transformed as
		\begin{eqnarray}
			\hat{\tilde{a}}(t) &\equiv& \hat T (x_c(t)) \hat a \hat T^\dagger (x_c(t))\\
			&=& \displaystyle e^{-i\frac{x_c(t) \hat p}{\hbar}}\sqrt{\frac{m\omega}{2\hbar}}(\hat{x}  + \frac{i}{m\omega}\hat{p} )e^{i\frac{x_c(t) \hat p}{\hbar}} \\
			&=& \sqrt{\frac{m\omega}{2\hbar}}(\hat{x} -x_c(t) + \frac{i}{m\omega}\hat{p} ) \\
			&=&\hat a -\sqrt{\frac{m\omega}{2\hbar}}x_c(t),
		\end{eqnarray}
		and the Hamiltonian part takes the form introduced in Eq.~\eqref{eq:hamiltonian} of the main text:
		\begin{eqnarray}
			\hat{H}(t) & = &\hat T(x_c(t)) \hat H_0 \hat T^\dagger(x_c(t)) \\
			& = &\frac{\hat p^2}{2m} + \frac{1}{2}m\omega^2 [ \hat x - x_c(t)]^2.\nonumber
		\end{eqnarray}
		Here, we used Campbell's identity,
		\begin{align}
			e^{s\hat \alpha } \hat \beta  e^{-s\hat \alpha} = \hat \beta  + s [\hat \alpha, \hat \beta], \end{align} when $[\hat \alpha, \hat \beta]$ is a c-number, and $s \in \mathbb{R}$ is a real number.
		
		Then, the quantum Lindblad master equation is written as 
		\begin{eqnarray}
			\label{eq:q-Lindblad}
			\dot{\hat{ \rho }}(t) & = & -\frac{i}{\hbar}[ \hat H(t),\hat\rho(t)] \\
			&+&\!\nu_- \left[ \hat{\tilde{a}}(t) \, \hat{\rho}(t) \, \hat{\tilde{a}}^{\dagger}(t) - \frac{1}{2} \left\{ \hat{\tilde{a}}^{\dagger}(t) \hat{\tilde{a}}(t), \hat{\rho}(t)\right\} \right] \nonumber   \\
			& +&\!\nu_+ \left[ \hat{\tilde{a}}^{\dagger}(t) \, \hat{\rho}(t) \, \hat{\tilde{a}}(t) - \frac{1}{2} \left\{ \hat{\tilde{a}}(t)\hat{\tilde{a}}^{\dagger}(t) , \hat{\rho}(t)\right\} \right] \nonumber \\
			& = & -\frac{i}{\hbar}[ \hat H(t),\hat\rho(t)] + {\mathcal{D}}_{L}.
		\end{eqnarray}
		We note that the Hamiltonian in terms of $\hat{\tilde{a}}$ reads
		is \begin{align}
			\hat H(t) = \hbar \omega \left[\hat{\tilde{a}}^\dagger\hat{\tilde{a}} +\frac{1}{2}\right],
		\end{align} which clarifies the role of Lindblad terms in Eq.~\eqref{eq:q-Lindblad} as excitations and de-excitations of the quantum harmonic oscillator centered at $x_c(t)$.
		
		For the Agarwal dissipator ${\mathcal{D}}_A$ \eqref{eq:Agarwal}, we have:
		\begin{eqnarray}
			& &\hat T(x_c) [\hat x,\{\hat p,\hat\rho\}]\hat T^\dagger(x_c) =
			[\hat x-x_c,\{\hat p,\hat\rho\}] \nonumber \\
			& = &[\hat x,\{\hat p,\hat\rho\}] - [x_c,\{\hat p,\hat\rho\}] = [\hat x,\{\hat p,\hat\rho\}].
		\end{eqnarray}
		where the commutator $[x_c,\cdot]=0$ vanishes since $x_c$ is a c-number. Analogously for the second term of the dissipator,
		\begin{eqnarray}
			& &\hat T(x_c) [\hat x,[\hat x,\hat\rho]]\hat T^\dagger(x_c)
			=
			[\hat x-x_c,[\hat x-x_c,\hat\rho]]\nonumber \\
			& = & [\hat x-x_c,[\hat x,\hat\rho]] = [\hat x,[\hat x,\hat\rho]],
		\end{eqnarray}
		all terms containing $x_c$ are commutators with a c-number, which therefore vanish. Thus, the Agarwal dissipator preserves its form under $\hat T(x_c)$; hence, in contrast with the Lindblad dissipator, adjustment to the actual position $x_c$ of the harmonic oscillator is not required.
		
		\section{Differentiability of the driving protocol}
		\label{sec:activation}
		
		Consider the von Neumann equation for the density matrix $\rho$:  
		\begin{equation}
			\partial_t \hat{\rho} = \frac{1}{i\hbar}[\hat{H}, \hat{\rho}].
		\end{equation}
		
		We perform a change of reference frame defined by the unitary translation operator $\hat{T}(x_c)$ \eqref{eq:transl},
		\begin{equation}
			\hat{\rho} = \hat{T}(x_c) \hat{\varrho} \hat{T}^\dagger(x_c),
		\end{equation}
		where $\hat{\varrho}$ is the density matrix in the new reference frame. Substituting this into the von Neumann equation yields  
		\begin{equation}
			\partial_t \hat{\varrho} = \frac{1}{i\hbar} [\hat{H}', \hat{\varrho}],
		\end{equation}
		with the transformed Hamiltonian  
		\begin{equation}
			\hat{H}' = \hat{H} - i\hbar \hat{T}^\dagger(x_c) \partial_t \hat{T}(x_c).
		\end{equation}
		
		For the translation operator, the derivative can be computed explicitly as  
		\begin{equation}
			\partial_t \hat{T}(x_c) = -\frac{i}{\hbar} (\partial_t x_c) \hat{p} \hat{T}(x_c),
		\end{equation}
		so that the transformed Hamiltonian becomes  
		\begin{equation}
			\hat{H}' = \hat{H} + (\partial_t x_c)\hat{p},
		\end{equation}
		requiring differentiability of the driving protocol $x_c$.
		
		\section{Details of the numerical procedure}
		\label{sec:simulation}
		For a given trajectory, $x_c(t)$ dictates the position of the minimum of a confining harmonic potential.
		After each $\delta t=10^{-3}$ a prediction for $\hat{\rho}(t+\delta t)$ is first calculated:
		\begin{equation}
			\hat{\rho}_{\rm pred}(t+\delta t)=\hat{\rho}(t)-\frac{i}{\hbar}[\hat{H}(x_c(t)),\hat{\rho}(t)] \delta t + {\mathcal{D}}(\hat\rho)\delta t. 
		\end{equation}
		Subsequently, this prediction $\hat{\rho}_{\rm pred}$ is combined with $\hat{\rho}(t)$ in the corrector step:
				\begin{equation}
			\hat{\rho}_m=\frac{1}{2}(\hat{\rho}(t)+\hat{\rho}_{\rm pred}(t+\delta t))
		\end{equation}
		\begin{equation}
			\hat{\rho}(t+\delta t)=\hat{\rho}(t)-\frac{i}{\hbar}[\hat{H}(x_c(t)),\hat{\rho}_m(t)] \delta t +  {\mathcal{D}}(\hat\rho_m)\delta t
		\end{equation}
		To obtain the mean squared displacement (MSD) at $t$ along a single given trajectory $x_c(t)$, we calculate:
		\begin{equation}
			\textrm{MSD}[x_c] = \textrm{Tr}\left({\hat{\rho}(t)\hat{x}^2}\right) - \textrm{Tr}\left({\hat{\rho}_{t = 0}\hat{x}^2}\right)
		\end{equation}
		which is then subsequently averaged over 200 stochastic trajectories, which has been demonstrated to be sufficient in \cite{Antonov/etal:2025}. The stochastic trajectories are calculated using the Euler-Maruyama scheme,
			\begin{subequations}
		\begin{eqnarray}
			\dot{x}_c(t+\delta t) & = &\dot{x}_c(t)\left(1 - \frac{\delta t}{\tau}\right) + \frac{\sqrt{2 D \delta} t}{\tau}\,\mathcal{N}(t),\\
			x_c(t+\delta t) & = & x_c(t) + \delta t\, \dot{x}_c(t+\delta t),
		\end{eqnarray}
	\end{subequations}
where $\mathcal{N}$ is a random Gaussian-distributed number. In numerical simulations, the density matrix is expressed in the Fock basis $\{ |n\rangle \}_{n=0}^{N-1}$, where $N$ is the Hilbert space truncation level:
\begin{figure}[htp!]
\vspace{-3ex}
\includegraphics[width=0.9\linewidth]{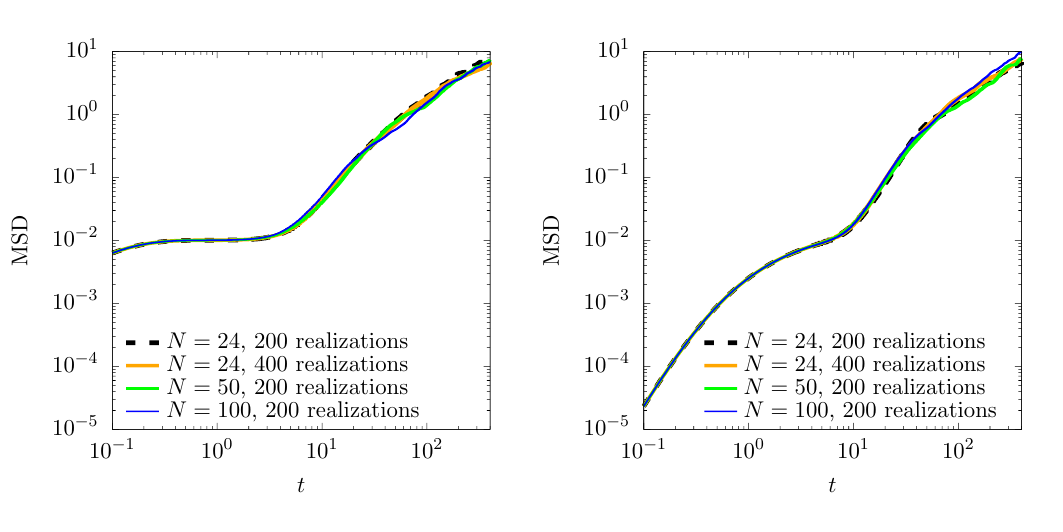}
\vspace{-4ex}
\caption{MSD of a quantum-active particle with (a) Lindblad \eqref{eq:Lindblad-dynamics} and (b) Agarwal \eqref{eq:Agarwal} dissipators for strong dissipation $\nu_+=10^{-1},\, \nu_-=10^1$ for various numbers of stochastic trajectories and Fock truncation. The thermalization rate is $\gamma = 2(\nu_--\nu_+)$ \eqref{eq:diss}, and the temperature is $ T = \hbar\omega/\left(k_B \ln\left[\nu_-/\nu_+\right]\right)$ \eqref{eq:temp}.
}
\label{Fig:8}
\end{figure}
		\begin{equation}
			\hat{\rho} = \sum_{m,n=0}^{N-1} \rho_{mn} |m\rangle \langle n |.
		\end{equation}
		In our simulations, we take $N = 24$ to balance computational cost and accuracy. In Fig.~\ref{Fig:8}, we check that further increasing the number of stochastic trajectories (ensemble size) or Fock basis does not affect the results. Yet, the Fock truncation may lead to violation of positiveness of the density matrix for the Agarwal dissipator, as will be shown below.
		
		To diagnose the physicality of the density matrix, we check the following parameters: (i) maximum deviation of $\Tr(\hat\rho)$ from one in a sample of trajectories $\max_{\textrm{traj}}\limits \{|1 - \Tr(\hat\rho)|\}$, (ii) minimum eigenvalue $\lambda_0$ of density matrix $\hat\rho$ in a sample of trajectories (or alternatively, $\max_{\textrm{traj}}\limits \{- \lambda_0\}$), and (iii) the average purity of the density matrix $\overline{\Tr(\hat{\rho}^2)}$ as a physical parameter. For the Lindblad dissipator in Fig.~\ref{Fig:9}(a), deviations of $\Tr(\hat\rho)$ from unity remain very small and converge to a stable plateau at late times, with no evidence of systematic drift or loss of numerical stability. The eigenvalues $\lambda_0$ are negative only within numerical errors; hence, the dynamics preserve the physicality of the density matrix. This behavior is not observed for the Agarwal dissipator with the same Fock truncation $N=24$, since the eigenvalues for some stochastic realizations quickly become negative, and trace preservation starts to break down near $t = 200$. This non-physicality is also reflected in the density matrix purity for the Agarwal dissipator in Fig.~\ref{Fig:9}(d), where it becomes larger than one, indicating the loss of physicality in the system. The appearance of negative eigenvalues is a consequence of the truncation of the Fock basis; the behavior becomes increasingly consistent as the truncation is increased, up to $N=100$ (Fig.~\ref{Fig:9}(b)), so it is reasonable to expect that no positivity violation would be observed in the ideal case of an infinite Fock basis. At the same time, it is clear that under identical numerical conditions with a finite Fock truncation, the master equation with the Agarwal dissipator is generally less stable than its Lindblad counterpart. From a physical perspective, for $N=100$ both the Lindblad and Agarwal dissipators exhibit consistent behavior: the density matrix purity $\overline{\Tr(\hat\rho^2)}$ starts at unity (pure quantum ground state) and quickly approaches $\overline{\Tr(\hat\rho^2)} \approx 0.98$ at later times, revealing that the system remains close to a pure quantum state.
			
		\begin{figure}[htp!]
			\vspace{-3ex}
			\includegraphics[width=0.9\linewidth]{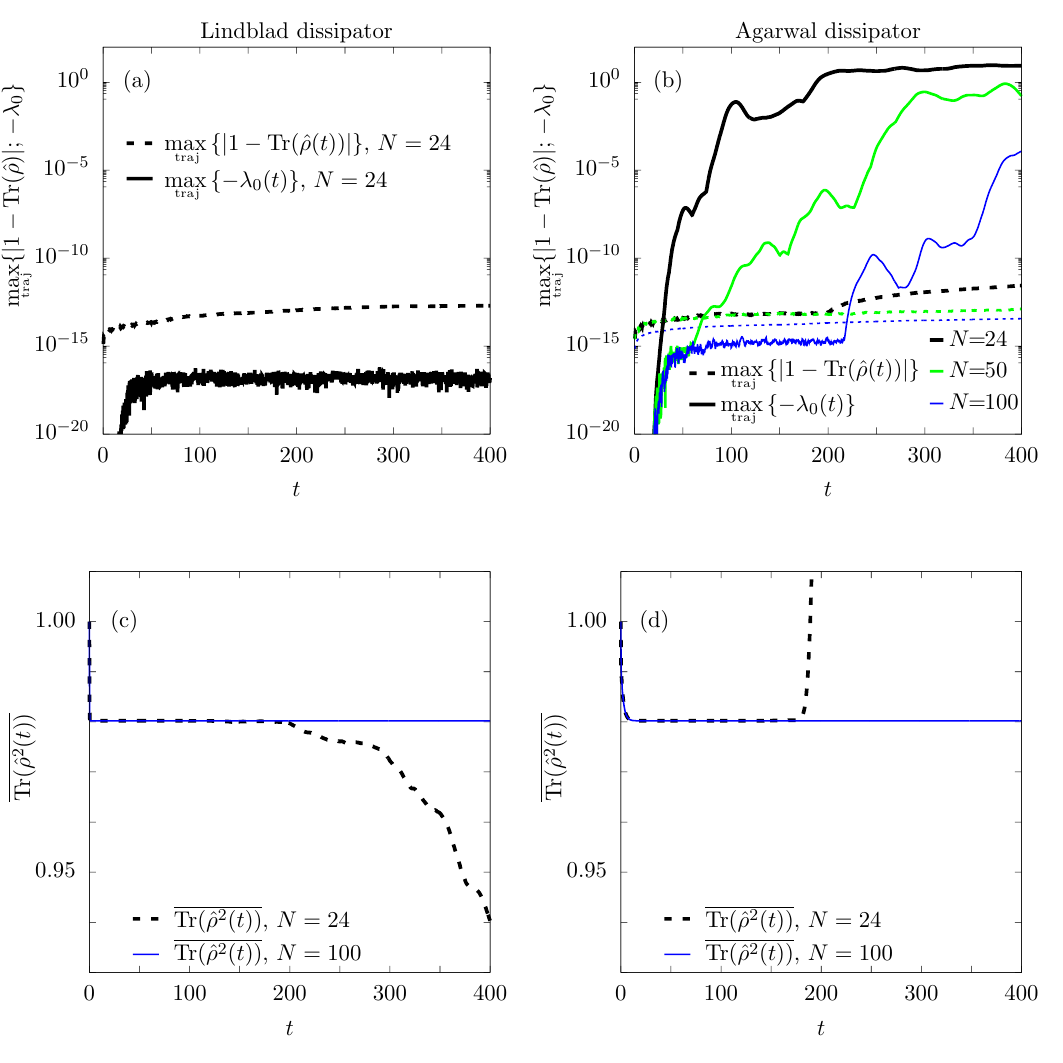}
			\vspace{-4ex}
			\caption{(a) Maximum deviation of $\Tr(\hat\rho)$ from one in a sample of 200 trajectories $\max_{\textrm{traj}}\limits \{|1 - \Tr(\hat\rho)|\}$ (dashed line) and minimum eigenvalue $\lambda_0$ of density matrix $\hat\rho$ in a sample of trajectories $\max_{\textrm{traj}}\limits \{- \lambda_0\}$ (solid line) for the Lindblad dissipator \eqref{eq:Lindblad-dynamics}, with $N=24$. (b) Same for the Agarwal dissipator \eqref{eq:Agarwal} with $N=24$ (thick black lines), $N=50$ (medium green lines), $N=100$ (thin blue lines). (c) Purity of the density matrix $\overline{\Tr(\hat\rho^2)}$ averaged over 200 stochastic trajectories for the Lindblad dissipator \eqref{eq:Lindblad-dynamics}, with $N=24$ (dashed black line) and $N=100$ (solid blue line). (d) Same for the Agarwal dissipator \eqref{eq:Agarwal}. The dissipation is strong with $\nu_+ = 10^1,\, \nu_- = 10^{-1}$, the thermalization rate is $\gamma = 2(\nu_--\nu_+)$ \eqref{eq:diss}, and the temperature is $T = \hbar\omega/\left(k_B \ln\left[\nu_-/\nu_+\right]\right)$ \eqref{eq:temp}.
			}
			\label{Fig:9}
		\end{figure}
			
		\end{widetext}
	\end{document}